# Schrödinger and Planck oscillators: not quite the same physics

**Enrique N. Miranda**[(1)(2)]

(1)Facultad de Ciencias Exactas y Naturales, Universidad Nacional de Cuyo, Mendoza, Argentina
(2)IANIGLA, CONICET, CCT Mendoza, Argentina

E-mail: emiranda@mendoza-conicet.gov.ar

**Abstract:**

In the statistical mechanics of quantum harmonic oscillators, the zero-point energy can either be included (Schrödinger oscillators) or omitted (Planck oscillators). For the usual results, the type of oscillator makes no difference but, looking more closely, it turns out that including or not this energy is not without consequences.

The chemical potential $\mu s$ of a Schrödinger oscillator set is calculated in the canonical formalism and this shows there is a temperature $T0$ for which $\mu s=0$; below this temperature, $\mu s>0$. When Planck oscillators are used instead, the chemical potential $\mu p$ is negative for all temperatures.

If the problem is approached in phonon terms and the system is considered to be in contact with a reservoir of particles (conditions of the grand canonical ensemble), a sort of critical temperature $Tc$ is found, for which the number of particles in the system diverges. For Schrödinger oscillators with $\mu s=0$, it turns out that $Tc=T0$, i.e. $T0$ is a reminder, in the canonical ensemble, of the divergent behavior in the number of particles when under the conditions of the grand canonical ensemble.

Also, a modified Einstein solid (MES) model is introduced. In this model the frequency of the oscillators changes with the volume of the solid, and this change is characterized by a certain value of the Grüneisen parameter. The bulk modulus of this solid can be calculated using Planck oscillators, and it becomes negative for certain temperature and volume values, which is physically incorrect. When Schrödinger oscillators are used, the bulk modulus is always positive. Therefore, the different behavior of both types of oscillators would indicate that only Schrödinger oscillators lead to correct results.

**I. Introduction**

The harmonic oscillator is ubiquitous in physics: it pops up in all the branches of the discipline, and is consequently a recurring topic in the teaching of both classical and quantum mechanics. It is in relation to the quantum harmonic oscillator that the concept of "zero-point energy" appears [1]. The zero-point energy is usually overlooked and often omitted in the calculations on the argument that it merely changes the starting point from which energy is measured and that it has no physical consequences. This is often the case, but occasionally it can be relevant; in quantum electrodynamics, for example, it is responsible for the Cassimir effect [2]. In this article, we will focus on statistical mechanics, and we will show that considering or not the zero-point energy can have consequences in some cases. After Pathria [3], we will call Schrödinger oscillators those which include the zero-point energy, and Planck oscillators those which omit it. In particular, we will show that the chemical potential behaves differently in both cases, and this could have interesting implications. Also, we will analyze an improved version of the Einstein solid, in which considering or not the zero-point energy is clearly important.

The structure of this article is detailed next. Section II considers a set of oscillators and evaluates the chemical potential in the canonical ensemble. Section III studies the Einstein solid and discusses the question regarding the chemical potential of phonons when working under the conditions of the canonical ensemble (fixed number of particles) and of the grand canonical ensemble (fixed chemical potential). Section IV introduces a modified version of the Einstein solid in which the frequency of the oscillators depends on the volume. The bulk modulus of this solid is evaluated and it is shown to differ according to the type of oscillator used. Finally, section V summarizes and discusses the results.

**II. A set of oscillators in the canonical ensemble**

The energy of a Schrödinger oscillator with frequency $\omega$ is $E_n = ½ \hbar\omega + n\hbar\omega$, where $n$ is a positive integer that indicates the quantum state of the system; as usual $\hbar = h / 2\pi$, with $h$ being the Planck constant. The first term is the so-called zero-point energy which is the topic of this article.

If $k$ is used to denote the Boltzmann constant, the partition function $zs$ will be:

$$zs = \sum_{n=0}^{\infty} e^{-E_n/kT} = e^{-\hbar\omega/2kT} \sum_{n=0}^{\infty} \left(e^{-\hbar\omega/kT}\right)^n$$

$$= \frac{e^{-\hbar\omega/2kT}}{1 - e^{-\hbar\omega/kT}}$$

(1)

In the case of a Planck oscillator, the zero-point energy is omitted, so that $E_n = n\hbar\omega$. Thus, the partition function $zp$ is:

$$zp = \sum_{n=0}^{\infty} e^{-E_n/kT} = \sum_{n=0}^{\infty} \left(e^{-\hbar\omega/kT}\right)^n$$

$$= \frac{1}{1 - e^{-\hbar\omega/kT}}$$

(2)

From these expressions, if we have a system formed by $N$ oscillators, the partition function of the set will be $Zp = (zp)^N$ for Planck oscillators, and $Zs = (zs)^N$ for Schrödinger oscillators. We should remember here that the Gibbs energy $G$ [4] in terms of the partition function is:

$$G = -kT\left(\ln Z - V\left(\frac{\partial \ln Z}{\partial V}\right)_T\right)$$

(3)

But since this model does not take into account the volume, the second term of (3) is null, and consequently we get:

$$Gp = kTN \ln\left[1 - e^{-\frac{\hbar\omega}{kT}}\right]$$

$$Gs = kTN \ln\left[1 - e^{-\frac{\hbar\omega}{kT}}\right] + N\frac{\hbar\omega}{2}$$

(4)

If we remember that the chemical potential is defined as [4]: $\mu = (\partial G / \partial N)_{T,P}$, it will be $\mu p$ for a set of Planck oscillators and $\mu s$ for Schrödinger oscillators. It is convenient to introduce here a characteristic temperature for the system, which is given by $\theta = \hbar\omega / k$. Thus, after some algebraic manipulations we get:

$$\mu p = kT \ln\left[e^{\frac{\theta}{T}} - 1\right] - k\theta$$

$$\mu s = kT \ln\left[e^{\frac{\theta}{T}} - 1\right] - \frac{1}{2}k\theta$$

(5)

Figure 1 shows the behavior of both potentials as a function of the temperature for $\theta = 1$.

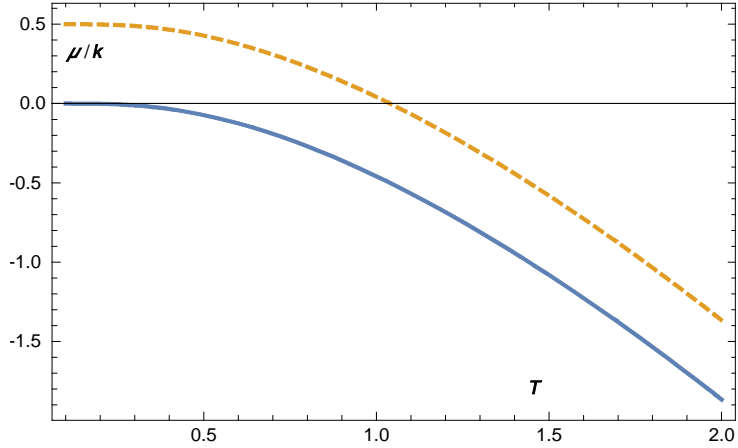

**Figure 1:** The chemical potential $\mu / k$ for a set of Planck oscillators (solid line) and a set of Schrödinger oscillators (dashed line) as a function of the temperature $T$. The characteristic temperature of the system is $\theta=1$. The qualitative behavior is different for both types of oscillators. For Planck oscillators, the chemical potential is always negative, while for Schrödinger oscillators there is a peculiar temperature where the chemical potential is zero, and below that, it becomes positive.

The qualitative difference in the behavior of Planck and Schrödinger oscillators is clearly seen. The chemical potential is always negative with Planck oscillators, while Schrödinger oscillators show an area where it is positive. And it is simple to find that temperature $T0$ where the chemical potential is null for Schrödinger oscillators. From the last line of (5), we get:

$$T0 = \theta/\left(2 \ln[2/(\sqrt{5} - 1)]\right) \cong 1.04\,\theta$$

(6)

The existence of this temperature is intriguing. Indeed, the chemical potential can also be defined in terms of the internal energy $U$ of a system that exchanges particles with the medium. Thus, we get [4]: $\mu = (\partial U / \partial N)_{S,V}$, that is to say, the chemical potential indicates how much the energy of the system changes by adding or removing a particle. If the chemical potential is null, this means that particles can be added with no cost in energy. A recent article [5] analyzed the behavior of a system formed by few Schrödinger oscillators in the microcanonical ensemble and found a temperature that depended on the size of the system and for which the chemical potential became null, exactly like in the present article. The authors interpreted that result as a sign of a Bose-Einstein condensation is a few-particle system. Although this interpretation is questionable, that paper agrees with what is stated in the present work:

there is a temperature for which the chemical potential of a Schrödinger oscillator set becomes null.

In the next section, the Einstein solid will be analyzed and the results will be reinterpreted in phonon terms.

### III. Einstein Solid

Einstein introduced the simplest model for solids that included characteristics of quantum mechanics [3, 6, 7]. It is assumed that $N_A$ atoms are fixed in a crystal lattice, and each can oscillate independently in the three spatial directions with a characteristic frequency $\omega$. Thus, we consider here a set of $3N_A$ oscillators that share the same frequency and we follow the same conditions of the previous section.

If we use Planck or Schrödinger oscillators, the chemical potential will be given by the expressions $\mu_P$ and $\mu_S$ presented in (5). However, there is an interesting conceptual difference. The Einstein solid was initially analyzed as a set of oscillators, but after the introduction of the phonon concept [3, 6, 7], its results were interpreted in terms of the mechanical waves and their associated pseudo-particles (the phonons). Therefore, the chemical potential corresponds to those particles, and if it becomes zero, then the chemical potential of the phonons becomes zero. In the case of this solid, the characteristic temperature is the so-called Einstein temperature $\theta_E$ and from (6) it can be inferred that $T_0 = 1.04\,\theta_E$. This means that for the Einstein solid, if Schrödinger oscillators are used, the chemical potential of the phonons is positive for $T < 1.04\,\theta_E$. In contrast, if Planck oscillators are used, the chemical potential is negative along the entire range of temperatures.

When reinterpreting the chemical potential of the oscillators in phonon terms, we encounter a problem. Indeed, conventional wisdom [3] states that phonons, as well as photons, are characterized by a chemical potential that is always negative. However, Figure 1 shows clearly that for $T < 1.04\theta$ the chemical potential is positive. How can we explain this contradiction? The answer is that we are working under different conditions. The result obtained in section II corresponds to a situation in which the number of oscillators is constant, and it should therefore be analyzed following the canonical ensemble. Conversely, the assertion regarding the negativity of the chemical potential of phonons and photons comes from the grand canonical ensemble, and corresponds to a situation in which the number of objects (oscillators, phonons, particles) is variable. To verify there are no contradictory results, we will analyze the Schrödinger and Planck oscillator sets in the framework of the grand canonical ensemble, that is, assuming constant $T$ and $\mu$ but fluctuating number of particles $n$.

The grand partition function $\Xi$ can be evaluated from the partition function of a particle $z_1$ [3, 7] with the formula:

$$\Xi = \frac{1}{1 - e^{\mu/(kT)} z_1}$$

(7)

Simply by replacing expressions (1) and (2) in (7), we can obtain the grand partition function of Schrödiger oscillators $\Xi s$ and of Planck oscillators $\Xi p$:

$$\Xi s = \frac{1 - e^{-\frac{\theta}{T}}}{1 - e^{-\frac{\theta}{T}} - e^{-\frac{\theta}{2T} + \frac{\mu}{kT}}}$$

$$\Xi p = \frac{1 - e^{-\theta/T}}{1 - e^{-\theta/T} - e^{\mu/(kT)}}$$

(8)

And once the grand partition function is known, the calculation of the average number *n* of particles is given by $n = (\partial(kT \ln \Xi)/\partial \mu)_T$ [3, 7]. From this, we get *ns* and *np* depending on whether we are working with Schrödinger or Planck oscillators:

$$ns = \frac{e^{\frac{\mu}{kT} - \frac{\theta}{2T}}}{1 - e^{-\frac{\theta}{T}} - e^{\frac{\mu}{kT} - \frac{\theta}{2T}}}$$

$$np = \frac{e^{\frac{\mu}{kT}}}{1 - e^{-\theta/T} - e^{\frac{\mu}{kT}}}$$

(9)

It is very interesting to graph *ns* and *np* for different values of the chemical potential but, before doing so, it is convenient to introduce the dimensionless variables *mu* and *t*, defined as:

$$\mu = mu \; k \; \theta$$

$$T = t \; \theta$$

(10)

Thus, the formulae (9) can be expressed in terms of the dimensionless variables given by (10). In addition, since the number of particles diverges under certain conditions, it is convenient to graph the inverse of the number of particles, i.e. *1 / ns* and *1 / np*. Figure 2a (2b) shows the inverse of the average number of particles for Schrödinger (Planck) oscillators in relation to the dimensionless temperature *t* and for different values of the dimensionless chemical potential *mu*. Obviously, it only makes physical sense when the average number of particles is positive. On the other hand, when the curves reach zero at a certain temperature, something odd happens: the average number of particles becomes infinite. Could this be interpreted as a Bose-Einstein condensation occurring for those temperatures? Maybe; but trying to answer that question would take us away from the goal of the present work. In any case, we will call *tc* the temperature at which the inverse of the number of particles equals zero, or to put it differently, when the average number of particles in the system diverges. In physical units, this is *Tc = tc θ*.

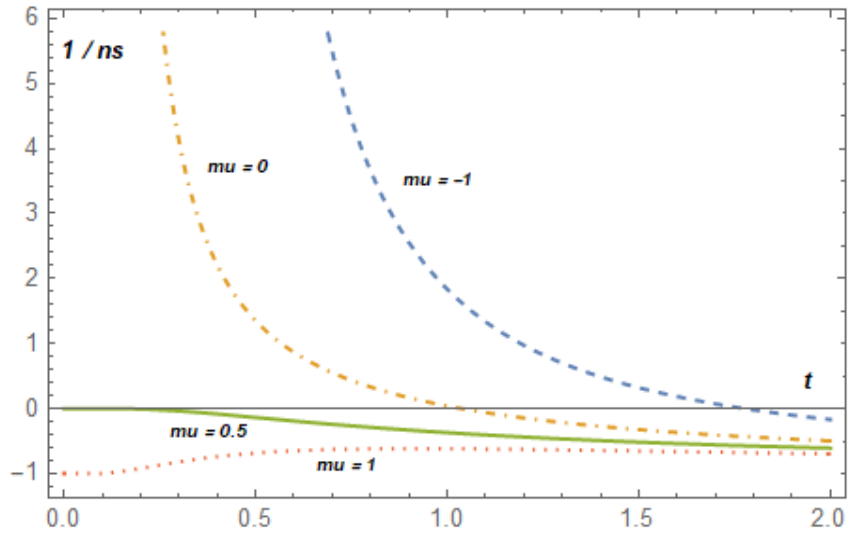

(a)

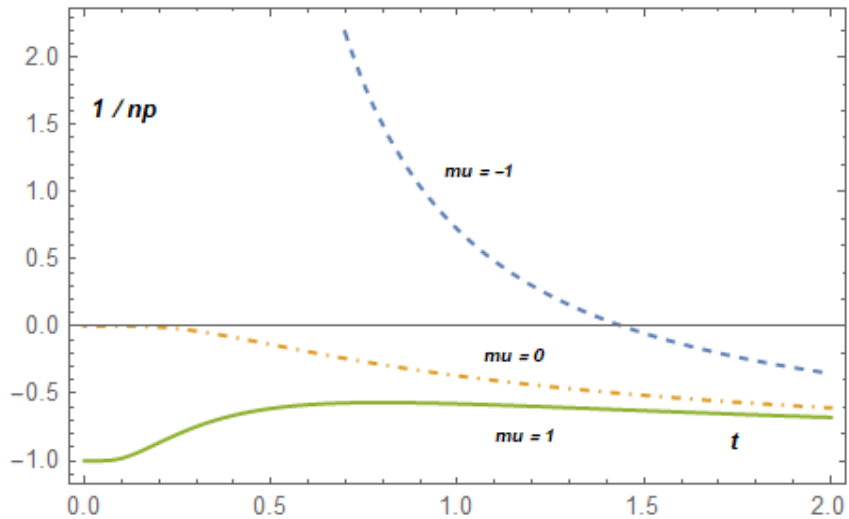

(b)

**Figure 2:** The inverse of the average number of particles (phonons) for Schrödinger (**a**) and Planck (**b**) oscillators as a function of the dimensionless temperature *t*. The different curves correspond to different values of the dimensionless chemical potential *mu*. Note that there is a temperature that we call *tc* and where the inverse of the number of particles equals zero; this means that the number of particles present in the system diverges. And this *tc* depends on the *mu*. For Schrödinger oscillators, *tc > 0* while *mu < ½*. When *mu = ½*, the behavior is different: there is a region of very low temperatures where the number of particles diverges. For *mu > ½*, the average number of particles is negative in the entire temperature range, which is physically absurd; this means that the chemical potential cannot surpass the value

½. When Planck oscillators are used, *tc* > 0 if *mu* < 0. It is not possible that *mu* > 0 since it leads to a negative average number of particles. The odd behavior is observed for *mu* = 0 with Planck oscillators and *mu* = ½ with Schrödinger oscillators.

The temperature *tc* exhibits an interesting behavior. Figure 2a shows, for Schrödinger oscillators, that *tc* > 0 if *mu* < ½. It is not possible that *mu* > ½ since it leads to a physically absurd result (a negative number of particles). And the behavior for *mu* = ½ is different: there is a region of very low temperatures where the number of particles of the system diverges. The same happens for Planck oscillators (Figure 2b), except that the divergence occurs for *mu* = 0. Here we can see again the relevance of considering or not the zero-point energy.

In the case of the Einstein solid, the number of oscillators (or phonons) is constant and, therefore, it should be treated within the canonical ensemble. To enable comparisons with the results of this section, we need to express the chemical potential $\mu$ in terms of the number of particles *ns* or *np*. From (9), we get:

$$\mu s = kT \left( \ln \left[ (\frac{ns}{ns+1})(e^{\frac{\theta}{T}} - 1) \right] \right) - \frac{1}{2} k\theta$$

$$\mu p = kT \ln \left[ (\frac{np}{np+1})(e^{\frac{\theta}{T}} - 1) \right] - k\theta$$

(11)

It must be noted that going from (9) to (11) is more than a simple algebraic manipulation. In (9) the chemical potential $\mu$ is known, the system is in contact with a reservoir of particles, and that determines the average number of particles *n*. In (11) the situation is the opposite: the number of particles is fixed, and that determines the chemical potential. Is it clear that in the thermodynamic limit, when the number of particles *n* is very big, *n / (n+1) → 1* and Eq. (5) is recovered. Thus, we can see that the assertion that the chemical potential of phonons has to be negative (or less than ½ for Schrödinger oscillators) does not contradict the result shown in Figure 1. The two cases simply deal with different physical conditions: in the first case, the system is in contact with a reservoir of particles while, in the other, the number of particles is fixed.

It is relevant to analyze what happens at the peculiar temperature *T0* observed in Figure 1 and corresponding to the temperature when $\mu s$ = 0. From (6) we know that *T0 = 1.04 θ*. And if we compare with the curve of *mu* = 0 in Figure 2a, we see that it is exactly for *T0* when the number of particles present in the system diverges. This means that for that temperature and chemical potential, if the system were in contact with a reservoir, it would incorporate an infinite number of particles. *T0* seems to be a sign in the canonical ensemble of a sort of "condensation" that would occur if one worked under the characteristic conditions of the grand canonical ensemble. And it is clear that this only appears if the zero-point energy is included in the energy of the oscillators, that is, it is not the same to work with Schrödinger or Planck oscillators.

In the next section, an improved version of the Einstein solid will be introduced and we will show that working with Planck oscillators leads to physically incorrect results.

**IV. Modified Einstein solid (MES)**

Another way to show the effect of the zero-point energy is to generalize the Einstein solid. For this, we will assume that the frequency of the oscillators depends on the volume, but first let us remember some definitions.

The Grüneisen parameter $\gamma$ [6, 8] can be defined in various ways [9], but the definition that better suits our purposes is the microscopic definition:

$$\gamma = -\frac{V}{\omega}\frac{\partial \omega}{\partial V}$$

(8)

That is, the Grüneisen parameter $\gamma$ is a measure of the change in the oscillator frequency $\omega$ when the volume $V$ of the solid changes. Here we can introduce a model, which we will call "modified Einstein solid" (MES), where the frequency is a function of the volume $\omega(V)$ and which is characterized by a certain value of the parameter $\gamma$. This means it is assumed that $\omega = C\, V^{\,\gamma}$, where $C$ is a constant that can be evaluated easily. If $\omega_0$ is the frequency of the solid when it is subjected to atmospheric pressure and its volume is $V_0$, therefore:

$$\omega = \omega_0 \left(\frac{V}{V_0}\right)^{-\gamma}$$

(9)

This modification to the Einstein solid has no effect whatsoever in the specific heat, but allows calculating the bulk modulus $K$ given by:

$$K = -V\left(\frac{\partial P}{\partial V}\right)_T$$

(10)

And if we remember that the pressure can be evaluated in terms of the partition function $Z$ as: $P = k_B T\, (\partial \ln Z / \partial V)_T$, it turns out that:

$$K = -k_B T V \left(\frac{\partial^2 (\ln Z)}{\partial V^2}\right)_T$$

(11)

At this point, we find that there are two different bulk moduli: *Kp* for Planck oscillators and *Ks* for Schrödinger oscillators:

$$Kp = -3\, N_A k_B T V \left(\frac{\partial^2 (\ln zp)}{\partial V^2}\right)_T$$

$$Ks = -3\, N_A k_B T V \left(\frac{\partial^2 (\ln zs)}{\partial V^2}\right)_T$$

(12)

All is left to do now is to replace the expressions *zp* and *zs* in (12) and to make the calculation. For keeping results simple, dimensionless variables are introduced and they are defined as: $t = T / \theta_E = Tk / \hbar\omega_0$ and $v = V / V_0$. This gives:

$$Kp = \frac{3 N_A k_B \theta_E}{V_0} \left( \frac{e^{\frac{v^{-\gamma}}{t}} v^{-1-\gamma}(1+\gamma)\gamma}{\left(e^{\frac{v^{-\gamma}}{t}} - 1\right)^2} - \frac{v^{-1-\gamma}(1+\gamma)\gamma}{\left(e^{\frac{v^{-\gamma}}{t}} - 1\right)^2} - \frac{e^{\frac{v^{-\gamma}}{t}} v^{-1-2\gamma}\gamma^2}{\left(e^{\frac{v^{-\gamma}}{t}} - 1\right)^2 t} \right)$$

$$Ks = \frac{3 N_A k_B \theta_E}{V_0} \left( \frac{e^{\frac{2v^{-\gamma}}{t}} v^{-1-\gamma}(1+\gamma)\gamma}{\mathbf{2}\left(e^{\frac{v^{-\gamma}}{t}} - 1\right)^2} - \frac{v^{-1-\gamma}(1+\gamma)\gamma}{\mathbf{2}\left(e^{\frac{v^{-\gamma}}{t}} - 1\right)^2} - \frac{e^{\frac{v^{-\gamma}}{t}} v^{-1-2\gamma}\gamma^2}{\left(e^{\frac{v^{-\gamma}}{t}} - 1\right)^2 t} \right)$$

(13)

These two expressions are clearly different, as can be appreciated by the highlighting (bold typeface) of the factors "**2**" that appear in the *Ks* formula but are absent in *Kp*. This means that, when a volume-dependent frequency is introduced, the use of Plank or Schrödinger oscillators leads to different results which can be observed quite easily because measuring the bulk modulus of a substance is a routine experiment. It should be noted that $V_0 / N_A$ is simply the specific volume per atom under normal pressure conditions, which is a known value, and so is the temperature $\theta_E$. Therefore, it is convenient to define a dimensionless bulk modulus as: $kp = Kp\, V_0 / (N_A\, k_B \theta_E)$ and $ks = Ks\, V_0 / (N_A\, k_B \theta_E)$. Figure 3 shows *kp* (solid line) and *ks* (dashed line) as a function of the dimensionless temperature *t* (Figure 3a) and of the dimensionless volume *v* (Figure 3b) using a Grüneisen parameter $\gamma = 2$, which is a typical value.

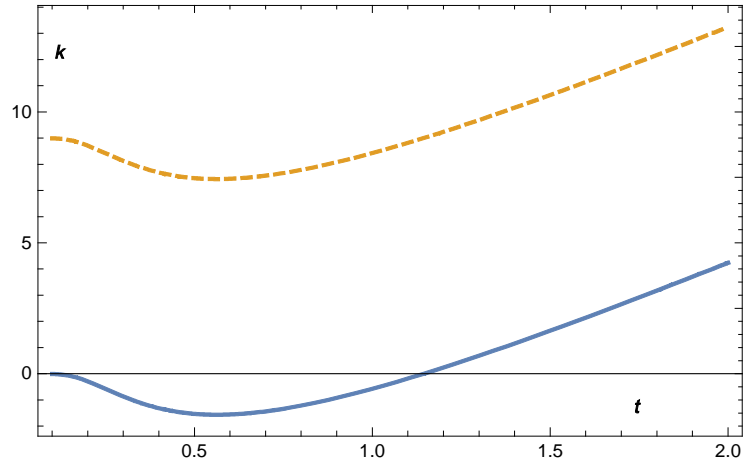

**(a)**

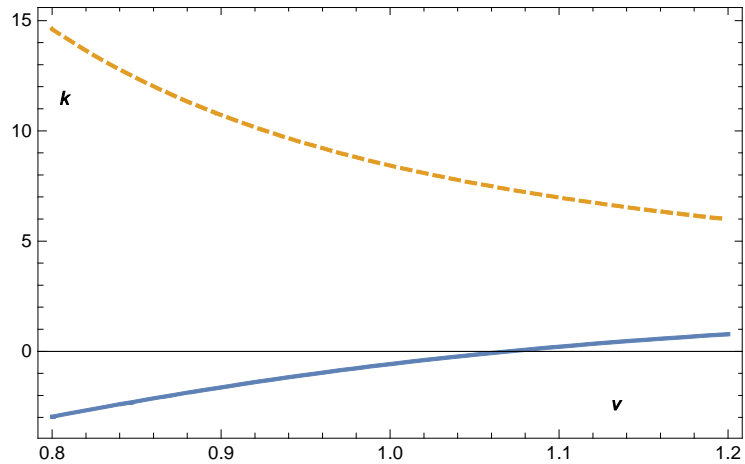

**(b)**

**Figure 3:** The dimensionless bulk modulus *k* as a function of the dimensionless temperature *t***(a)** and of the dimensionless volume *v***(b)**. The solid line is obtained from Planck oscillators, while the dashed line corresponds to Schrödinger oscillators. As is a typical value, a Grüneisen parameter $\gamma = 2$ was used. In (a) the case for a fixed volume $v = 1$ is considered, and in (b) for a fixed temperature $t = 1$. Note in both cases that *k* becomes negative for Planck oscillators, and that is physically incorrect. The result that makes physical sense is obtained with Schrödinger oscillators.

The significant outcome from Figure 3 is that the bulk modulus becomes negative when Planck oscillators are used, and this is physically impossible. From this it can be inferred that the use of Planck oscillators leads to incorrect results. Instead, with Schrödinger oscillators, the bulk modulus results positive in the entire range of temperatures, as should be.

At this point we may wonder what happens with the chemical potential in the MES. To answer this question, we need to go back to (3) and repeat the calculations, which have become more complicated because of the explicit dependence of the partition function *Z* on the volume *V*. This can be a good exercise for students and we will omit the details. We will only mention that the chemical potential in the MES is positive for a temperature $T \lesssim 6\, \theta_E$ when working with Planck oscillators, and $T \lesssim 7.5\, \theta_E$ when working with Schrödinger oscillators. This means that the chemical potential is positive with both types of oscillators and in the entire temperature range where the model is valid.

**IV. Conclusions**

The intention in this article has been to show the physical implications of taking into account or not the zero-point energy of an oscillator. The term ½ $\hbar\omega$ in the energy of a quantum oscillator is not irrelevant. And we have shown this in two cases.

On the one hand, we consider a set of oscillators and evaluate the chemical potential. Results are different when Planck or Schrödinger oscillators are used. In the first case, the chemical potential is negative in the entire temperature range, while in the second, there is a temperature *T0* below which the chemical potential is positive. The significance of such a temperature where the chemical potential becomes zero is unclear; however, analyzing the system in terms of the grand canonical ensemble can shed some light on the matter. Within this ensemble, it is verified that the chemical potential $\mu$ has to be lower than ½$\hbar\omega$ for Schrödinger oscillators and lower than zero for Planck oscillators to keep the number of particles positive. In addition, there is a sort of critical temperature *Tc* for which the number of particles in the system diverges. This critical temperature depends on the chemical potential and is zero for $\mu$ = ½$\hbar\omega$ (Schrödinger oscillators) or $\mu$ = 0 (Planck oscillators). In the case of Schrödinger oscillators, *Tc* = *T0* is verified for $\mu$ = 0; this means that the peculiar temperature *T0* is a remainder, in the canonical ensemble, of the divergence in the number of particles when working under the conditions of the grand canonical ensemble.

On the other hand, we consider a modified Einstein solid (MES), characterized by a certain temperature $\theta_E$ and a certain value for the Grüneisen parameter $\gamma$, i.e. this is a solid where the frequency of the oscillators changes with the volume. This case shows very clearly that it is not the same to use Planck or Schrödinger oscillators. Using Planck oscillators to evaluate the bulk modulus leads to negative results for certain volume and temperature values, and that is physically impossible. With Schrödinger oscillators, the bulk modulus is positive for all temperatures and volumes. Regarding the chemical potential, it is positive in the entire validity range of the model.

Finally, it should be noted that some statistical mechanics textbooks omit the zero-point energy when dealing with the harmonic oscillator and the Einstein solid, both in the canonical and the microcanonical ensembles. And surely, many university lecturers (the present author included) also omit it. But in this article, such a practice has been proven incorrect. Although there are no differences in the common results, when we look into subtler questions, differences do arise. Certainly, Schrödinger and Planck oscillators do not lead to the exact same physics.